\documentclass[%
 english,
 reprint,
 superscriptaddress,
 amsmath,amssymb,
 aps,
]{revtex4-1}

\usepackage{graphicx}% Include figure files
\usepackage{dcolumn}% Align table columns on decimal point
\usepackage{bm}% bold math
\usepackage[english]{babel}
\begin{document}

%\preprint{APS/123-QED}

\title{Anomalous dynamics of interstitial dopants in soft crystals}% Force line breaks with \\

\author{Justin Tauber}%
\thanks{J.T. and R.H. contributed equally to this work.}
\affiliation{%
 Physical Chemistry and Soft Matter, Wageningen University \& Research, Stippeneng 4, 6708 WE Wageningen, The Netherlands
}%

\author{Ruben Higler}%
\thanks{J.T. and R.H. contributed equally to this work.}
\affiliation{%
 Physical Chemistry and Soft Matter, Wageningen University \& Research, Stippeneng 4, 6708 WE Wageningen, The Netherlands
}%

\author{Joris Sprakel}%
 \email{joris.sprakel@wur.nl}
\affiliation{%
 Physical Chemistry and Soft Matter, Wageningen University \& Research, Stippeneng 4, 6708 WE Wageningen, The Netherlands
}%

\date{\today}% It is always \today, today,

\begin{abstract}
The dynamics of interstitial dopants governs the properties of a wide variety of doped crystalline materials. To describe the hopping dynamics of such interstitial impurities, classical approaches often assume that dopant particles do not interact and travel through a static potential energy landscape. Here we show, using computer simulations, how these assumptions and the resulting predictions from classical Eyring-type theories break down in entropically-stabilised BCC crystals due to the thermal excitations of the crystalline matrix. Deviations are particularly severe close to melting where the lattice becomes weak and dopant dynamics exhibit strongly localised and heterogeneous dynamics. We attribute these anomalies to the failure of both assumptions underlying the classical description: i) the instantaneous potential field experienced by dopants becomes largely disordered due to thermal fluctuations and ii) elastic interactions cause strong dopant-dopant interactions even at low doping fractions. These results illustrate how describing non-classical dopant dynamics requires taking the effective disordered potential energy landscape of strongly excited crystals and dopant-dopant interactions into account. 
\end{abstract}

\maketitle

Doping pure crystalline solids  with small amounts of interstitial impurities is a widely used method to enhance material properties such as heat and electric conductivity\cite{Pei2013, Gaume2003, Kenyon2002, Macfarlane1999} or to tailor mechanical properties\cite{Hentschel2015}. Prototypical examples include the introduction of carbon atoms in iron crystals to make steel or the doping of plastic crystals with Li-ions to create solid-state batteries\cite{Macfarlane1999}. To ensure longevity of doped materials, it is essential that the spatial homogeneity and transport dynamics of the dopants within the crystal are well controlled and understood. Although theories and models are abundant\cite{Homan1964, Farraro1979, Weller2006, Tapasa2007, Xiao2015, Kang2015, Hentschel2015}, it remains unclear how large thermal excitations of the matrix lattice affect the dynamics of dopants. This becomes of particular interest during the processing of doped crystals, where they are heated close to or beyond their melting point. For example in body-centered cubic (BCC) iron doped with carbon, significant deviations from the exponential increase of diffusivity with temperature, expected from Arrhenius' law, are observed close to the melting temperature where lattice excitations are strong\cite{DaSilva1976}. While doping is typically performed to tailor material properties at the macroscopic scale, these enhanced properties emerge from the dynamics and interactions between dopants at scale of individual atoms \cite{Ramamoorthy1996}. In classical theories for dopant dynamics, impurity particles are described as hopping through a potential energy landscape which is set by a perfect lattice symmetry, with transition rates governed by the energy barriers between adjacent interstitial sites and their occupancy \cite{Wert1950, Homan1964, Farraro1979}. In reality, thermal fluctuations of atoms away from their equilibrium lattice positions will randomize the instantaneous potential energy landscape that the dopants experience; this could lead to failure of classical approaches to capture the physics of impurity diffusion when lattice excitations become pronounced. This may be particularly severe for crystals of the BCC symmetry, such as the high-temperature lattice of sodium,  lithium and iron. In these high-temperature BCC phases, thermal fluctuations are large due to the relatively low coordination number; in fact, these fluctuations increase the entropy of the solid to such an extent that they are responsible for its thermodynamic stability \cite{Haasen1992}. For impurity transport in structurally-disordered colloidal glasses, it was recently shown that thermal fluctuations which create time-variations in the potential energy landscape can have a strong effect on the dopant diffusivity \cite{Sentjabrskaja2016, Evers2013}; yet these effects remain largely unexplored for very soft crystals which exhibit an on-average ordered lattice. 

In this paper we study the dynamics of interstitial dopants in BCC crystals prepared from colloidal particles interacting by long-ranged electrostatic interactions. Using Brownian Dynamics simulations we probe in detail how strong thermal fluctuations of the base crystal affect the spatial homogeneity of the dopants and their motion through the lattice. Dopants within a static base crystal obey quantitative predictions of classical transition-state theory; by contrast, the same impurities diffusing in a fluctuating crystal exhibit completely different behaviour. We show how thermal excitation of the lattice causes clustering of the interstitials while simultaneously giving rise to strong disorder in the instantaneous potential energy landscape. This results in heterogeneous and anomalous dynamics of interstitials within an on-average perfect lattice. We support these observations with direct imaging experiments on a colloidal system using confocal microscopy. These data illustrate how large thermal fluctuations can give rise to heterogeneous dynamics in ordered solids, which cannot be captured by classical hopping theories. 

The classical approach to describe the diffusion of interstitial impurities through a crystalline matrix starts with the assumption that the dopants experience a static potential energy landscape set by the summation of interactions between a dopant and all particles in the base crystal \cite{Wert1950}. Assuming that interactions between dopants are negligible, i.e.~that the dopant concentration is low and the interstitial site occupancy approaches zero, this reduces to a simple transition-state theory for thermally-activated jumps between neighbouring minima in the energy landscape. 

In a BCC crystal the minima in which interstitial impurities will reside are the tetrahedral sites (Fig.~1B, green spheres) \cite{Wang2015}. We can identify two transition paths between tetrahedral sites that are most likely to contribute to the motion of a dopant. The first comprises the shortest path from one tetrahedral site to another (T--T transition) during which displacement the particle crosses a saddle point in the energy landscape. The second (T--O--T transition) goes from a tetrahedral site through an octahedral site to an adjacent  tetrahedral site \cite{Wang2015}. The rate at which these hops occur is governed by the energy barrier $U_A$ separating two sites along either path. 

\begin{figure}%[tbhp]
\centering
\includegraphics[width=\linewidth]{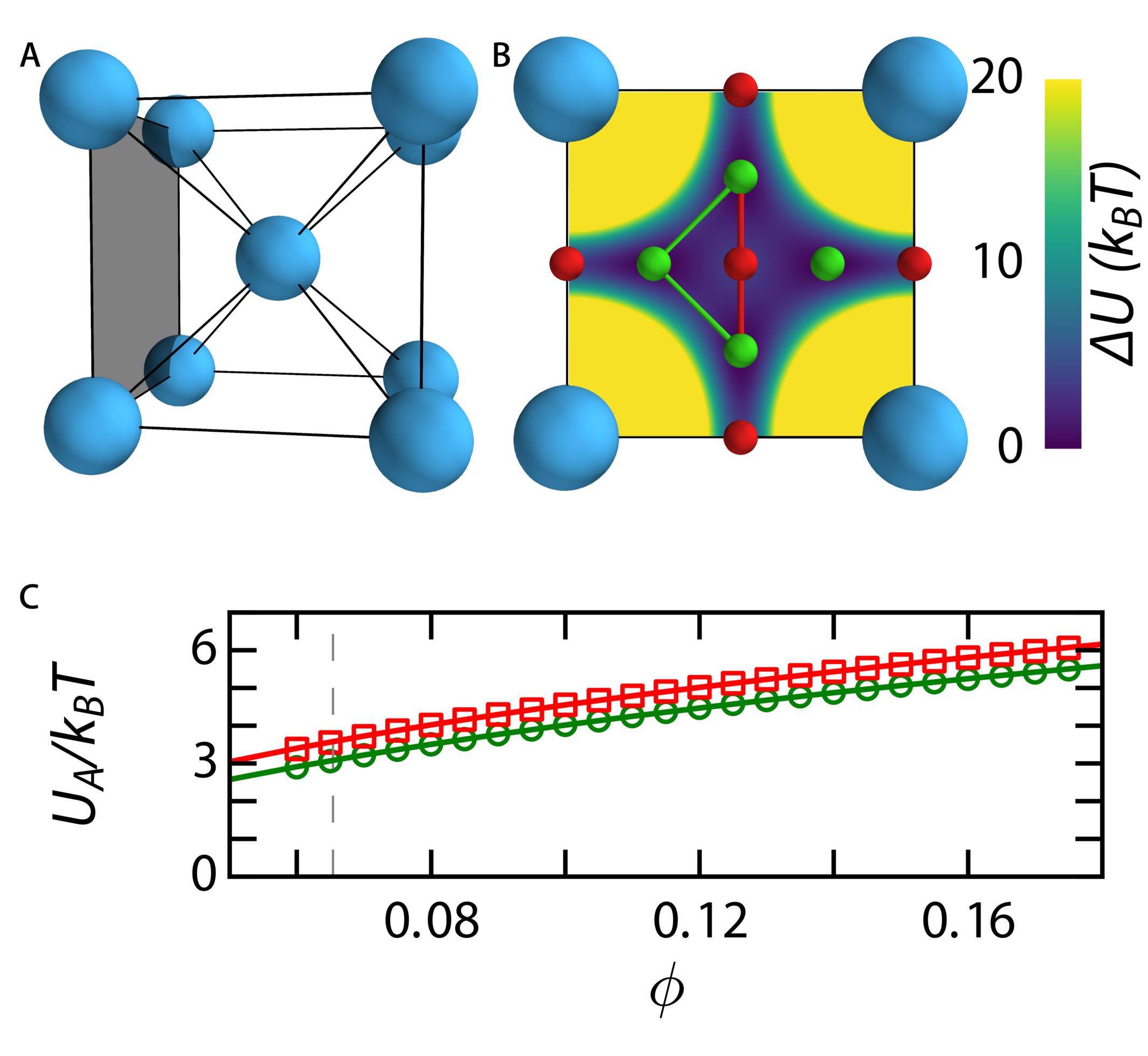}
\caption{A. Schematic representation of a BCC unit cell. B. The interstitial sites in a plane of the BCC unit cell with tetrahedral sites in green and octahedral sites in red. The T--T path (green) and T--O--T path (red) are indicated with lines connecting the interstitial sites. The potential field felt by the dopant is shown in the background. The values are the potential energy with respect to the global minimum at the tetrahedral site. Yellow indicates values of $20~k_BT$ and above. C. Hopping barrier $U_{A}(\phi)$ along a T--T (circles) and T--O--T path (squares) from numerical calculations. The solid lines are a parametric fit to $U_A(\phi)$ as described in the text, the dashed line indicates the melting point $\phi_m$.}
\label{fig:figure1}
\end{figure}

We parametrize our simulations to match an experimental system of charged PMMA particles in an apolar solvent, which forms BCC crystals at low densities \cite{Yethiraj2003, Kanai2015}. In these colloidal systems, the main control parameter is particle volume fraction $\phi$. The crystals are formed from colloids with a diameter $\sigma_b = 1.8 \mu$m and doped with interstitial impurities with $\sigma_d = 0.9 \mu$m. The interactions are described by Yukawa potentials to map the simulated phase behaviour as a function of $\phi$ onto that determined experimentally (see Fig.~S1 \& methods). For a perfect BCC lattice we can now compute the activation energy for both the T--T and T--O--T paths by summing the potential energy fields, taking long-ranged contributions into account. The BCC crystal exhibits a periodic network of energy  minima (see Fig.~S3), which provides an efficient means for interstitial motion on large length scales\cite{Wang2015}. For the colloidal BCC crystal, the numerically enumerated transition energies are few to several $k_BT$ and the difference in activation energy between the T--T and T--O--T paths are small (symbols in Fig.~1C). To describe these data phenomenologically, we consider the difference between the summed potential field at the interstitial site where $U$ exhibits a minimum and the transition maximum $U_{A} (\phi) = U_{+} (\phi) - U_{-} (\phi)  = \epsilon \left(\frac{e^{-\kappa g_{+}a(\phi)}}{g_{+}a(\phi)} -  \frac{e^{-\kappa g_{-}a(\phi)}}{g_{-}a(\phi)}\right)$, where  $a(\phi) = \left(\frac{\pi}{3\phi}\right)^{\frac{1}{3}}$ is the normalised lattice constant in units $\sigma_b$ and the geometrical constants $g_{-}$ and $g_{+}$ account for the potential energy fields at the minima and maxima, respectively. We use this empirical equation to fit the simulation data at discrete values of $\phi$; with values of $g_{+} = 0.348$; $g_{-} = 0.345$ for  the T--T transition and $g_{+} = 0.297$; $g_{-} = 0.295$ for the T--O--T this relation describes our numerical calculation data well (lines Fig.~1C).

Within the classical approach, the rate at which transitions occur is governed by a thermally-activated process of the Eyring type: $ k_{h} = k_{h,0} \exp({-{U_{A}}/{k_BT}})$. The Brownian attempt frequency is given by $k_{h,0}(\phi) = {D_{0}}/{d_{h}^{2}(\phi)} = {k_BT}/({d_{h}^{2}(\phi) 6 \pi \eta \sigma_{d} / 2})$, in which $d_{h}$ is the length of the transition path and $D_0$ is the self-diffusion coefficient of the interstitial impurities in a solvent of viscosity $\eta$. The long-time diffusion coefficient of the interstitial impurities as a function of volume fraction of the BCC crystal can now be predicted as  $D_{l}(\phi) = d_{h}^{2}(\phi) k_{h}(\phi) = D_{0} \exp ({-({U_{A}(\phi)}/{k_BT})^{\beta}})$ in which the stretch exponent $\beta$ accounts for a distribution in hopping times due to the similar barriers of the two different transition paths.

%figure2
\begin{figure*}[t]
\centering
\includegraphics[width=\linewidth]{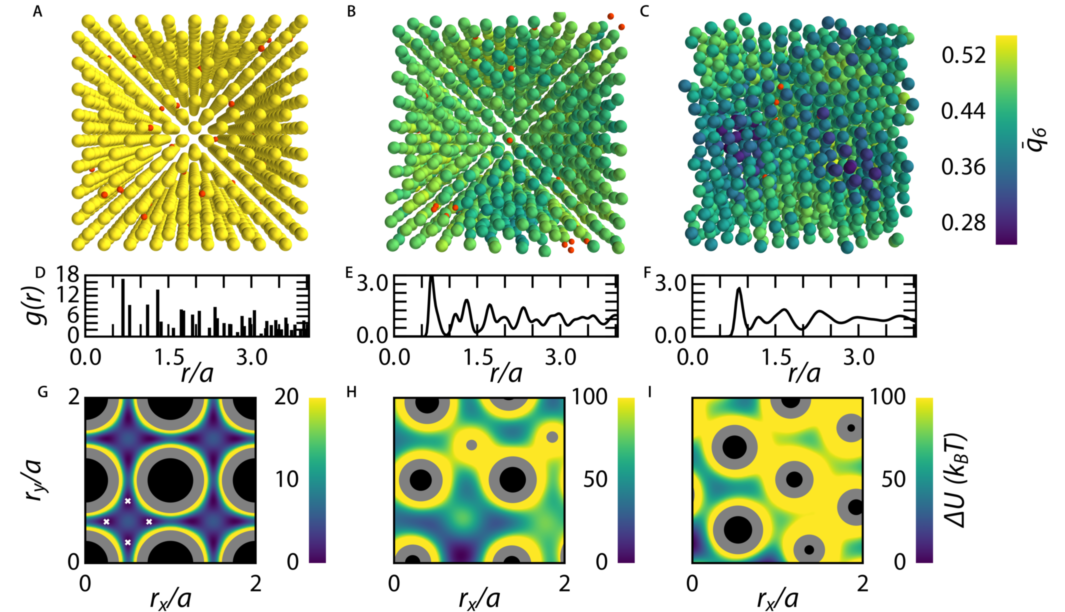}
\caption{Structural features of the BCC crystal for the simulation with a static (A,D and G) and dynamic base crystal (B, E and H) and the experiments (C, F and I). A-C: snapshots of BCC crystals at $\phi = 0.14$, in which particles are color-coded according to their instantaneous bond-order parameter $\bar{q_6}$ and dopant particles rendered in orange(A \& B only). D-F: Pair correlation functions $g(r)$. G-I: Potential energy landscape through four unit cells within the on-average ordered lattice, where black disks indicate the hard-sphere radius of the base crystal particles and the grey areas indicate the volume, from which the centre-of-mass of dopants are excluded, due to dopant-base particle overlap. }
\label{fig:figure2}
\end{figure*}

To test the validity of this prediction based on classical transition-state theory, we simulate the Brownian dynamics of interstitial impurities within a static and perfect BCC crystal (Fig.~2A). The potential energy field experienced by the dopants exhibits clear minima at the tetrahedral sites (crosses, bottom Fig.~2G). This leads to characteristic hopping dynamics in the trajectories of individual interstitial impurities, with particles vibrating within a tetrahedral site until they hop to a neighboring site (Fig.~3A\&C). Over time, the interstitial impurities probe the entire matrix by travelling through the interconnected network of local minima (see Fig.~S3). This gives rise to a mean-squared displacement $\langle\Delta r^2 (\tau)\rangle$ as shown in the left panel of Fig.~4B); at short times vibrations within the interstitial sites give rise to subdiffusive motion. This transitions into diffusive behavior at times longer than the Brownian self-diffusion time, $\tau \gg \tau_B$, as particles explore the lattice by hopping between interstitial sites; this is  characterised by a long-time diffusion coefficient $D_l$ (circles in Fig.~4A \& Fig.~S5). The simulation data for this static scenario are described very well by the prediction for $D_l(\phi)$ from transition-state theory, with $\beta = 0.61 \pm 0.01$ used as a fit parameter (line in Fig.~4). The fact that $\beta$ deviates from unity indicates a heterogeneous hopping process occurring both via the T--T and T--O--T transitions; the relative occurrence of T--T versus T--O--T hops is expected to be 3.5:1 based on the difference in activation energies taking into account the number of possible T--T and T--O--T transitions from a given tetrahedral site. We note that at this point, we have not established an exact and quantitative relationship between the value of $\beta$ and the ratio of hops occurring via the two possible transition routes. 

In real materials, at least one crucial assumption in this classical approach fails as the matrix in which dopants diffuse itself is also excited by thermal fluctuations. Especially for body-centered cubic crystals in close proximity to their melting point, where doped crystals are typically processed to induce ductility and malleability, these fluctuations are known to be strong \cite{Haasen1992}. Allowing the BCC phase in these colloidal systems to fluctuate retains an on-average perfect structure as evident from distinct Bragg peaks in their structure factor (see Fig.~S2). However, snapshots of the instantaneous structure show significant deviations from a perfect lattice as particles displace significantly from their equilibrium positions. Reconstructions of the system in which the particles are color-coded according to their instantaneous bond-order parameter $\bar{q_6}$\cite{Lechner2008} illustrate the significant amount of thermal disorder within these BCC crystals, both in-silico and in experiments (Fig.~2B\&C). The thermally-excited excursions of particles from their average lattice position translate into peak broadening in the pair-correlation function $g(r)$ (Fig.~2E\&F). We note that $g(r)$ for experiment and simulation are in excellent agreement, even though the field-of-view in our measurements is limited due to experimental constraints. Despite the strong thermal disorder in these fluctuating BCC crystals, it can still be structurally distinguished from a liquid by means of spherical harmonic bond-order parameters (see Fig.~S4), to probe local structure, and the existence of well-defined Bragg peaks in the structure factor  (Fig.~S2) which signals the presence of long-ranged order. 

The effect of the instantaneous deviations from a perfect lattice due to thermal excitations becomes apparent when we plot a snapshot of the potential energy landscape that a dopant particle experiences at a given time. Instead of the regular landscape which exhibits minima at tetrahedral sites, the fluctuating BCC crystal presents an apparently disordered potential energy landscape (Fig.~2H) in which the variations in the height of energy barriers and the depth of localisation wells are significantly larger as compared to the perfect lattice. Also from experimental data we can reconstruct the potential energy landscape; we obtain the particle positions from three-dimensional image stacks. Using the pair interaction potential obtained by inversion of pair correlation functions\cite{Behrens:2001kq} and assuming pairwise additivity, we can compute the potential energy of inserting a dopant particle at a given location within the lattice. Also the energy landscapes reconstructed in this way from snapshots of the experimental system, exhibit strong disorder (Fig.~2I). 

%figure 3
\begin{figure}%[tbhp]
\centering
\includegraphics[width=\linewidth]{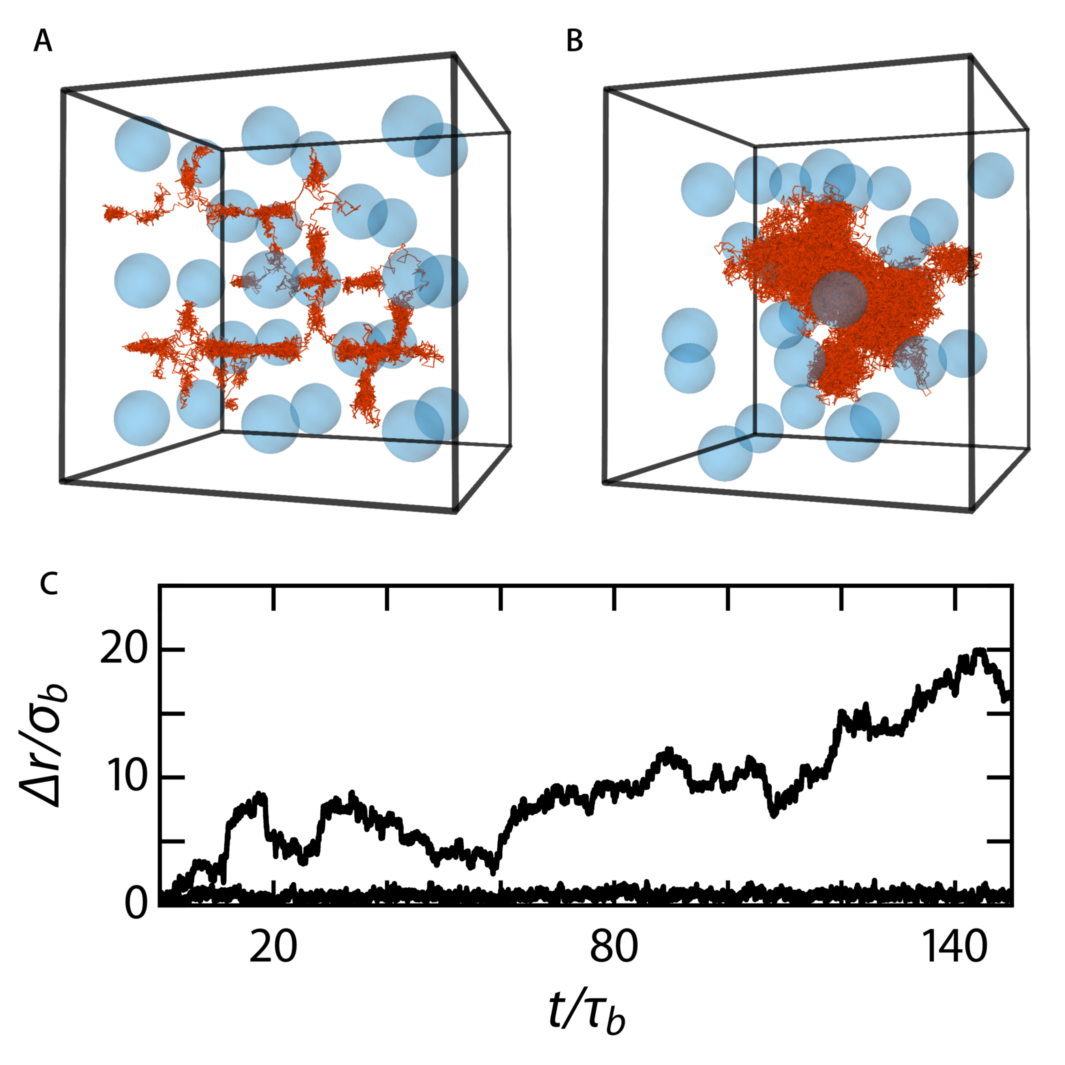}
\caption{A,B: Trajectory of a single interstitial dopant in the crystalline matrix over $\Delta t =$~29~$\tau_{b}$ in a static (A) and $\Delta t =$~150~$\tau_{b}$ in a dynamic BCC crystal (B). C: interstitial displacement with respect to $t=0$ in a static (bottom line) and dynamic base crystal (top line)}
\label{fig:figure3}
\end{figure}

This high degree of instantaneous disorder in the energy landscape results in very different interstitial dynamics than those predicted by the classical theory. The dopant particles are more strongly localised, and transitions between minima appear at much lower frequency as compared to a static crystal (Fig.~3). As a consequence, the ensemble-averaged mean-square displacements exhibit a localisation plateau which extends by several orders of magnitude (right panel, Fig.~4B), resulting in a strongly reduced rate of diffusion at long times. To confirm that the interstitial mean-squared displacement converges to a diffusive behaviour at long times, we run a longer simulation up to $2 \cdot 10^4~\tau_B$; indeed the upturn we see in Fig.~4B becomes diffusive at even longer times (Fig.~S5C).

To extract $D_{l}$ from these data, we extrapolate the mean-squared displacement to infinite time; see SI for a detailed description of our method. Allowing the crystal that surrounds the interstitial impurities to fluctuate results in more than two orders-of-magnitude reduction in the diffusion rate (blue symbols Fig.~4A). Clearly, the effect of thermal excitations of the lattice cannot be ignored in describing dopant dynamics in BCC crystals.

Two possible contributions to this drastic reduction in interstitial diffusion rate can be identified. First, static or low-temperature BCC crystals feature a percolated path of T-T transitions, providing an efficient pathway for interstitial diffusion over large length scales\cite{Wang2015}. This percolated path results from the centre-of-inversion symmetry of the BCC lattice. In the thermal BCC phase, especially close to melting, thermal excitations of the lattice are so pronounced that the \emph{instantaneous} centre-of-inversion symmetry is lost. Note that this only applies to instantaneous snapshots of the structure, whereas time-averaging cancels out these fluctuations and restores the BCC symmetry, for example evidenced by the distinct Bragg peaks in the time-averaged structure factor (Fig.~S2). As thermal fluctuations break the local and instantaneous symmetry, the percolated transition path that relies on this symmetry is also lost; this is evidenced in potential energy isosurfaces reconstructed from snapshots of the thermal BCC lattice in Fig.~S3. 

Secondly, as the potential energy landscape is strongly time-varying, hopping now requires not only a fluctuation large enough to escape a local minimum, but also the simultaneous availability of a low-energy pathway that remains open during the transition event. In effect, two competing frequencies come into play; i) that of escape attempts of the dopant and ii) the frequency with which the potential energy landscape reconfigures. As the Brownian time scales of the base crystal and the dopants do not differ by much due to the moderate size asymmetry, escape events now become cooperative and thus significantly less likely. It is known that the effect of fluctuating barriers on hopping is strongly non-monotonic and can lead to either enhancement, when the two frequencies become resonant, or reduction in transition rates \cite{Reimann1997,Schweizer2004}. As we work in the classical limit, where the transition itself is not instantaneous but requires a finite time, this poses the additional constraint that the path remains open for the duration of the transition event, which further slows down hopping. The combination of these events leads to a strong quenching of the interstitial mobility in fluctuations BCC lattices.

A key feature for particles in disordered potential landscapes is the emergence of heterogeneous dynamics. To investigate this, we plot the time-averaged $\langle\Delta r^2 \rangle$ for all interstitial particles individually. For the static crystal, no heterogeneities in particle dynamics are observed, with all mean-squared displacements collapsing onto the ensemble average (Fig.~4B left panel). By contrast, for the fluctuating BCC crystal, strongly heterogeneous dynamics are observed, with a large inhomogeneity in the single-particle behavior (Fig.~4B right panel). 

To explore the origins of these distinct heterogeneous dynamics within a on-average ordered solid, we reconstruct snapshots of the interstitial positions. While dopants are homogeneously distributed for the static crystal (Fig.~5A), they exhibit strong clustering in the fluctuating BCC over the entire range of base crystal densities $\phi$ (Fig.~5B and Fig.~S7). We hypothesise that this clustering is caused by the lattice strain accompanying the insertion of a single interstitial impurity into a tetrahedral site. Clustering between interstitials minimizes the overall elastic deformation of the matrix and is thus energetically favorable. This gives rise to an emergent elastic attraction between the impurity particles. Similar lattice-strain mediated interactions are well-established to exist for crystallographic defects that cause a lattice deformation\cite{Teodosiu:2013ui}. Indeed, we observe a strong increase in the lattice strain, defined as the average displacement of base crystal particles from their equilibrium position $\Delta r_{b,l}$ normalised to the lattice constant $a$, as a function of the distance to a nearest impurity.

We observe that the clusters are highly dynamic, with spontaneous particle association and dissociation (Fig.~S6-8 and Video~S9 \& 10). This indicates a dynamic equilibrium between singlets ($S$) and bound states ($B$) $nS \rightleftharpoons B_{n}$, in which the association constant depends on the effective attractive potential $U_{eff}$ emerging through the elasticity of the matrix: $k_a \propto \exp(U_{eff}/k_BT)$. We observe a significant fraction of singlets in stable coexistence with clusters, which does not evolve over time after equilibrating our simulation system (see Fig.~S9). This suggests that the effective attraction strength  is of the order of the thermal energy $k_BT$; the dynamic equilibrium between clusters and singlets resulting from a balance between the configurational entropy of distributing impurities across the lattice and the enthalpic gain upon forming a cluster. This is further corroborated by the distribution of cluster sizes $P(S_{C})$ (Fig.~5F). These data are well described by an exponential decay as indicated by the solid line in Fig.~5F. This indicates that clusters are formed by an open association process governed by a dynamic reaction equilibrium between unimeric dopants and clusters.

%figure4
\begin{figure}[h]
\centering
\includegraphics[width=\linewidth]{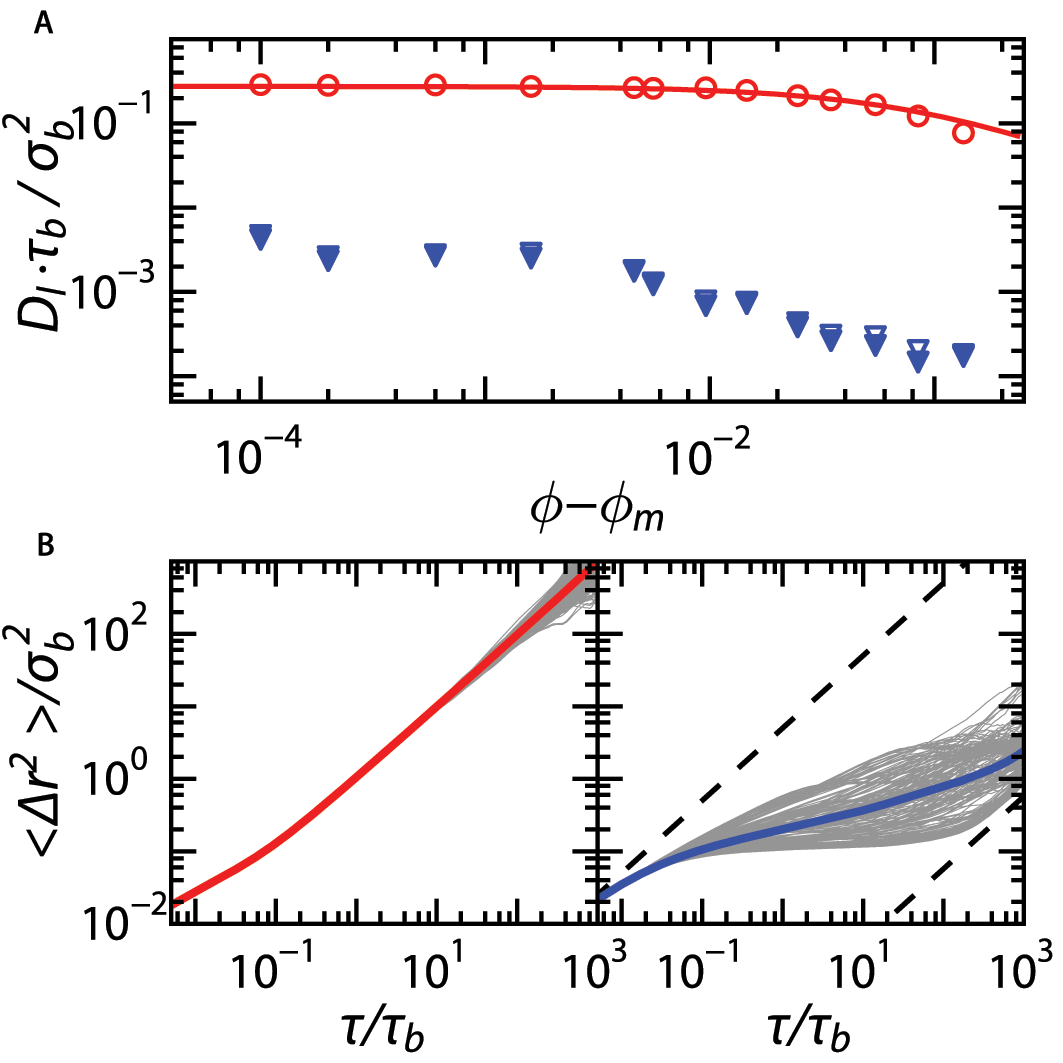}
\caption{A.  Long-time diffusion coefficients $D_l$ as a function of distance to the melting point $\phi - \phi_m$ for static (circles) and dynamic crystals (triangles), with $\phi_m = 0.061$ as determined in the SI Fig.~S1. Open symbols indicate $D_l$ determined from the mean-squared displacements at $\tau = 5 \cdot 10^2 \tau_b$, while filled symbols are computed by extrapolating $\langle\Delta r^{2}\rangle$ to infinity. Solid line is a fit to the transition-state prediction for $D_l(\phi)$, as described in the text. B:  $\langle\Delta r^{2}\rangle$ for individual particles, with the ensemble-average $\langle\Delta r^{2}\rangle$ (thick line) superposed for fixed (left) and dynamic crystal (right). }
\label{fig:figure4}
\end{figure}

Intuitively, one may expect that particles present in an attractive cluster of dopants would exhibit lower mobility than their singlet counterparts as their local density is higher. Surprisingly, we observe the opposite; trajectories of dopants reveal that the degree of localisation is in fact reduced for particles in clusters as compared to singlets within the same lattice (Fig.~5D). To determine the origins of this counter-intuitive observation, we determine the instantaneous deviation of particle positions away from their equilibrium site in the lattice $\Delta r_{b,l}$ as a function of the distance to the nearest dopant $\Delta r_{b,d}/a$ . Especially for low volume fractions, where deviations from classical transition-state theory are large, we observe a strong increase in the lattice strain in proximity to a dopant (Fig.~5E) while the average orientational bond-order $\bar{q_6}$ is maintained (see Fig.~S4). This suggests that dopant particles, especially those present in clusters, locally weaken the lattice, resulting in larger mobility for both the dopants and the surrounding crystalline matrix. Interestingly, the fact that deviations in the dynamics of interstitial impurities are exacerbated close to the melting transition is also observed in the carbon-doped BCC phase of iron \cite{DaSilva1976}. 

\begin{figure}[t]
\centering
\includegraphics[width=\linewidth]{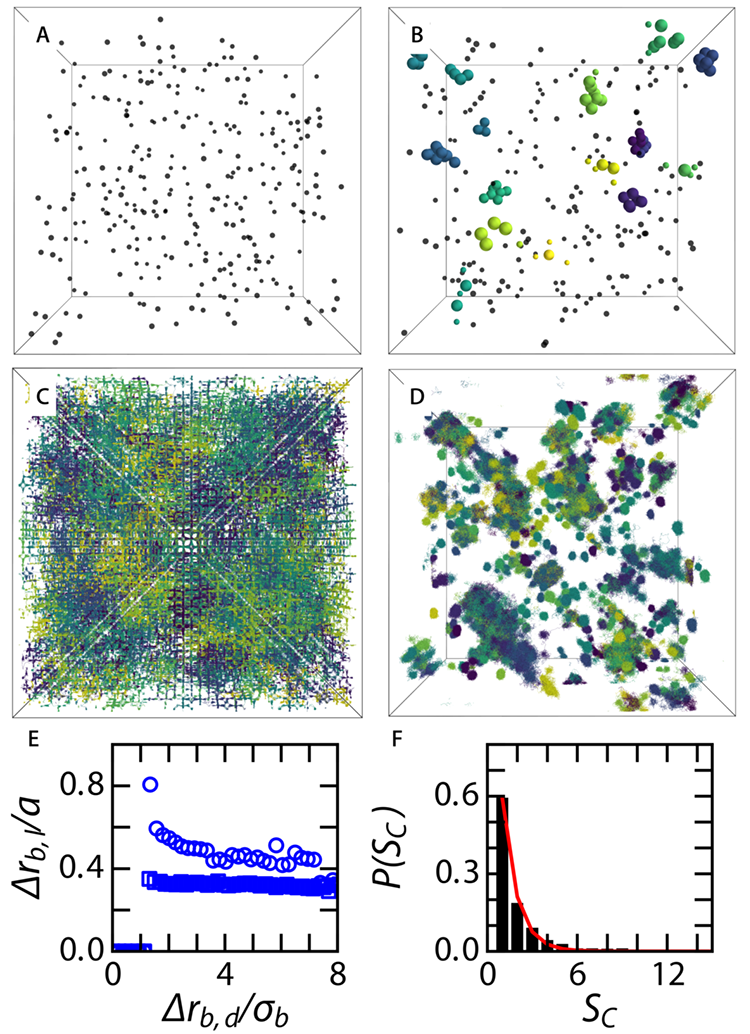}
\caption{A, B: Dopant positions in a snapshot at $\phi = 0.07$, in which particles belonging to the same cluster are color-matched, while singlets are displayed in gray for static (A) and dynamic base crystals (B). C,D: Particle trajectories of all particles over a time interval of $2 \cdot 10^3 \tau_b$ for static (C) and dynamic (D) base crystal. E: Lattice strain $\Delta r_{b,l}/a$, taken as the average deviation of particle positions with respect to their equilibrium site, as a function of distance to a dopant for $\phi = 0.07$ (circles) and $0.12$ (squares). F. Distribution of cluster sizes $P(S_{C})$ for $\phi = 0.07$, fitted with an exponential distribution $P(S_{C})\propto e^{-S_c/S_c^*}$ with $S_c^*= 0.96$ the characteristic cluster size(red line).}
\label{fig:figure5}
\end{figure}

In this paper we demonstrated how thermal fluctuations can lead to the failure of classical theories for dopant dynamics and gives rise to complex heterogeneous and anomalous dynamics within an on-average ordered matrix. Large instantaneous deviations from a perfect lattice due to thermal excitations cause a disordered potential energy landscape in which interstitial atom diffusion can be orders of magnitude slower than expected based on transition-state theory. Our simulations also give rise to a microscopic picture of the strongly heterogeneous dynamics of interstitial dopants: elastic interactions between dopants cause them to agglomerate within the lattice, which in turn locally soften the matrix and gives rise to enhanced mobility. The coupling between spatial organisation of the dopants, the local properties of the matrix and resulting dopant dynamics can be expected to play a crucial role in the effective tailoring of material properties using doping. Arriving at a complete description of these complex dynamics would require extension of classical lattice dynamics to account for both the fluctuating and locally disordered energy landscapes, for which a framework was developed for glasses \cite{Schweizer2004}, and for the emergent interactions and spatial inhomogeneity of the dopants.

\section*{Materials \& Methods}

\subsection*{Simulations}
We perform Brownian Dynamics simulations using HOOMD-BLUE, a GPU accelerated software package, in single-precision mode.\cite{Anderson2008, Glaser2015} Analysis routines are all written in the python programming language, using scipy,\cite{Jones2001} numpy\cite{VanDerWalt2011}, lmfit\cite{Newville2014}, scikit-learn\cite{Pedregosa2011}, matplotlib\cite{Hunter:2007}, and mayavi\cite{Ramachandran:2011} libraries.
For the calculation of bond order parameters we use BondOrderAnalysis\cite{Lechner2008} and we calculate Voronoi cells using the voro++ package\cite{Rycroft2009}. All quantities are expressed in normalised units, in terms of base particle diameter $\sigma_{b}$, base particle self-diffusion time $\tau_{b}$ and $k_BT$ respectively. All raw data and scripts can be accessed via https://github.com/sprakellab/dopantdynamics.

Simulations are performed in the canonical ensemble (or N,V,T-system) with periodic boundaries. Systems consist of two types of particles, one that forms the crystalline matrix ($\sigma_b = 1.8 \mu$m) and the dopants ($\sigma_d = 0.9 \mu$m). This size ratio of 0.5 is experimentally accessible and close to that for carbon-doped iron and lithium-impurities in silicon.\cite{Slater1964} The particles interact via Yukawa potentials parameterised using experimental data (see below). The simulations assume pairwise additivity of the potentials; in the experimental system of charged colloids many-body effects are known to occur\cite{Merrill:2009ab}. Nonetheless, in previous work we have established that pairwise additive BD simulations can capture the main behavior of experimental crystals of the charged colloids we simulate here\cite{Meer:2014ab}. 

Brownian Dynamics integration, using the overdamped Langevin equations, are performed with a time step of $2.5\cdot 10^{-5} \tau_b$, in one of two ways, either both particle types are integrated for the dynamic crystal or only the dopant particles are subjected to integration for the static matrix. In all cases we simulate $N = 13718$ base crystal particles. Dopant particles are placed randomly at tetrahedral interstitial sites in the pristine BCC lattice in a ratio of 1:47. Simulations are run for at least $2 \cdot 10^3 \tau_b$, preceded by an equilibration time of $2 \cdot 10^2 \tau_b$. 

\subsection*{Experiments}
Some aspects of the simulation results are experimentally verified by studying a system of polymethyl methacrylate particles, stabilised by polyhydroxystearic acid.\cite{Kanai2015}
Particles with diameters of $\sigma_{b} = 1.8~\mu m$ and $\sigma_{b} = 0.9~\mu m$ are prepared using established procedures\cite{Antl1986}.  We suspend the particles in a density matching solvent mixture of \textit{cis}-decalin and tetrachloroethylene, in which 10 mM Aerosol OT is added to charge the particles\cite{Kanai2015}. We image the samples in three-dimensions and time using confocal fluorescence microscopy using a VisiTech Infinity-3, mounted on a Nikon Ti-U and equipped with a Hamamatsu ORCA-Flash 4.0 camera. Three-dimensional volumes of 50 x 50 x 30 $\mu m^3$ are acquired at 1Hz. Particle centroid positions are determined and linked together in time using well established methods based on the fitting of a Gaussian curve\cite{Gao2009}.

\subsection*{Mapping}
The particles in the simulation interact via the Yukawa potential $U(r)/k_BT =  \epsilon \frac{\exp{(-\kappa \sigma (\frac{r}{\sigma} - 1))}}{r / \sigma}$. In the solvent we use, the inverse screening length $\kappa$ is determined to be $1.8/\sigma_{b}$ \cite{Kanai2015}. To define the interaction strength $\epsilon$ we map the simulation data onto the experimentally-determined melting point of the BCC crystal (Fig. S1). An $\epsilon_{b,b}$ of $713$ gives a melting point at a volume fraction $\phi$ of $0.061$ \textit{in silico} close to the melting point found experimentally, $\phi = 0.060$. For the smaller dopants we assume a particle-size independent surface charge density such that $\epsilon_{d,d} = 227$. The cross interactions between dopant and matrix is taken as the average of the base-base interaction and the dopant-dopant interaction ($\epsilon_{b,d} = 470$). Data analysis methods are described in the SI.

% Bibliography
\nocite{Ester1996}
\bibliography{dopant}

\begin{titlepage}
	\centering
	{\Huge Supplementary Information \par}
	\vspace{1.5cm}
	{\Huge\itshape Anomalous dynamics of interstitial dopants in soft crystals\par}
	\vspace{0.5cm}
	{\large\itshape Justin Tauber\textsuperscript{a,1},Ruben Higler\textsuperscript{a,1} \& Joris Sprakel\textsuperscript{a,2}\par}
	\vspace{0.5cm}

	{\normalsize \textsuperscript{a}Physical Chemistry and Soft Matter, Wageningen University \& Research, Stippeneng 4, 6708 WE Wageningen, The Netherlands}
	\vfill
	{\footnotesize\itshape\textsuperscript{1}J.T.(Justin Tauber) and R.H. (Ruben Higler) contributed equally to this work.\par}
	{\footnotesize\itshape\textsuperscript{2}To whom correspondence should be addressed. E-mail: joris.sprakel@wur.nl\par}
\end{titlepage}

\cleardoublepage
\twocolumngrid

\subsection{Figure S1: Determination of the phase diagram}
To accurately define the melting point $\phi_m$ and the region where liquid and BCC crystal coexist, we calculate, for every particle, the averaged bond order parameter following the approach described in [21]:

\begin{equation}
\bar{q}_l(i) = \sqrt{\frac{4 \pi}{2l + 1}\sum^l_{m = -l}{|\bar{q}_{lm}(i)|^2}}
\end{equation}

\noindent for $l = 6$ or $l = 4$ and where

\begin{equation}
\bar{q}_{lm} = \frac{1}{\widetilde{N}_b(i)} \sum_{k = 0}^{\widetilde{N}_b(i)}{q_{lm}(k)}
\end{equation}

\noindent where $N_b$ is the number of neighbours of particle $i$. We identify nearest-neighbors based on proximity, considering only particles closer together than the lattice constant $a$. The spherical harmonic bond order parameter $q_{lm}(k)$ for particle $i$ is given by

\begin{equation}
q_{lm}(i) = \frac{1}{N_b(i)} \sum_{j = 1}^{N_b(i)}{Y_{lm}(\mathbf{r}_{ij})}
\end{equation}

\noindent where $Y_{lm}$ are the spherical harmonics. 

Using $\bar{q}_6$ we characterize the local structure. We reconstruct our in-silico system by color-coding all particles according to their $\bar{q}_6$ value for the three different regimes: liquid, liquid-BCC coexistence, and solid BCC (Fig.~S1A-C top). This clearly shows the different regimes in our phase diagram. In order to identify the phase transition points we calculate the probability distribution $P(\bar{q}_6)$. Here we average over 90 snapshots with $dt = 2 \cdot 10^1$ $\tau_{b}$ (Fig.~S1A-C bottom).
We find three distinct behaviors.
In the liquid we observe is a sharp peak around a $\bar{q}_6$ value of $0.2$ (Fig.~S1A bottom). The solid BCC (including dopants) exhibits a peak around a value of $0.4$ with a tail extending towards lower $\bar{q}_6$ values, due to the dopant induced lattice strain (Fig.~S1C bottom).

In the phase coexistence regime we observe a peak at $\bar{q}_6 =0.4$, corresponding to the BCC crystal structure, and a large tail extending to lower values corresponding to the coexisting liquid pockets (Fig.~S1C bottom).

We locate the melting point, where liquid pockets first appear, at $\phi  \sim 0.0655$, and the freezing point, where the last remnants of solid vanish at $\phi  \sim 0.0605$. These transitions are in agreement with those found in the experimental system onto which our simulation parameterisation is mapped. 

\subsection{Figure S2: Calculation of two-dimensional structure factor}
We calculate a two-dimensional projection of the three-dimensional static structure factor $S(\mathbf{q})$ as:

\begin{equation}
S(\mathbf{q}) = \frac{1}{N}\left \langle \sum_{jk}{\exp{i \mathbf{q} \cdot (\mathbf{r}_j - \mathbf{r}_k)}} \right \rangle
\end{equation}

\noindent averaged over all particles and time (Fig.~S2). We find well defined Bragg peaks corresponding to a BCC crystal for the static base crystal, the dynamic base crystal, and the experimental BCC phase system. 

\subsection{Figure S3: Percolating network of tetrahedral interstitial sites}

To illustrate the percolated network of tetrahedral hopping transitions we plot an isosurface ($U(\mathbf{r}) = 3.5~k_BT$) of the calculated three-dimensional potential field for the static base crystal (Fig.~S3A). The interconnected transition path network can be clearly seen. By contrast, a similar analysis for the fluctuating dynamic base crystal ($U(\mathbf{r}) = 42~k_BT$), reveals a striking difference with a complete vanishing of the interconnected structure, and only localised blobs of lower potential energy remaining (Fig.~S3B). Both potential fields are calculated on simulation snapshots of the system on a grid of 100x100x100 voxels. 

\subsection{Figure S4: Influence of dopants on base crystal structure}

In order to determine the local structure of the BCC phase we plot the average bond parameters $\bar{q}_{4}$ and $\bar{q}_{6}$ for every base particle (blue) and every dopant (green), for both a static base crystal (Fig.~S4A left) and a dynamic base crystal (Fig.~S4A right). As a reference we have plotted $\bar{q}_{4}$ and $\bar{q}_{6}$ for particles in a perfect BCC lattice (red) and particles in a liquid state (yellow, obtained from simulations at $\phi=0.600$). 

The fact that the structure of the base crystal is influenced by thermal fluctuations is apparent from the $\bar{q}_6$ bond parameter probability distribution (Fig.~S1B). The addition of dopants results in the appearance of a tail in $P(\bar{q}_6)$ at $\bar{q}_6 < 0.46$, indicating that the dopants locally strain the BCC lattice (Fig.~S4B). 

\subsection{Figure S5: Determination of the long time diffusion coefficient}

The dynamics of the dopant particles are studied by means of their mean-squared displacement $\langle\Delta r^2\rangle$.
For every particle we calculate $\langle\Delta r^2\rangle$ from the dopant particle trajectories for $200$ values of $\tau$ logarithmically spaced over the entire simulation duration. 
In addition, we perform an ensemble-average over all particles, for both static (Fig.~S5A red curves) and dynamic (Fig.~S5A blue curves) base crystals. We define a effective diffusion rate as the local slope of the mean-squared displacement $D = \frac{1}{6} \frac{d\langle\Delta r^2\rangle}{d\tau}$ (Fig.~S5B). To extract the long time diffusion coefficient $D_l$ we use two methods: First we take $D_l$ to be the diffusion coefficient at $\tau = 5 \cdot 10^2 \tau_b$. Secondly we fit the tail-end of the $D(\tau)$ curve with the function $D(\tau) = -1 + c^{\frac{1}{\tau}} + D_{l}$, where both $c$ and $D_{l}$ are used as fit parameters. The values found for $D_l$ using both methods are in close agreement (Fig.~4A blue open \& closed triangles). To show that at long times the MSD goes towards diffusive behavior we performed an extended simulation. The mean-squared displacements calculated from this simulations clearly show the upturn at long times towards a slope of one; indicative of diffusive motion (Fig.~S5C). 

\subsection{Figure S6: Voronoi analysis}

To evaluate the local environment of dopant particles, we use a Voronoi analysis; we calculate Voronoi cells using voro++[32] taking into account the periodic boundaries of our simulation box. We observe a significant difference in the local environment of the dopant particles in the static base crystal compared to the dynamic base particles. In the case of a static base crystal the distribution in Voronoi volume is very narrow both averaged over time and in a simulation snapshot (Fig.~S6A-B red). The dynamic base crystal case shows a much larger distribution of Voronoi cell volumes (Fig.~S6A-B blue), again confirming the highly heterogeneous nature of the instantaneous structure of soft BCC crystals.
We also compare the number of faces of a Voronoi cell around a dopant for the cases of a static and a dynamic base crystal. For a dopant which resides at the tetrahedral site in a perfect BCC crystal the expected number of faces is $8$; this is indeed what we find in a static crystal (Fig.~S6C red), confirming that virtually all interstitial dopants reside in tetrahedral sites. The few cells which exhibit a higher number of faces correspond to dopants which are in transition between two tetrahedral sites since these data are taken from snapshots of the dopant structure. The situation in the dynamic crystal is significantly different, the loss of well defined interstitial sites causes the dopant particles to have a Voronoi cell faces distribution close to what we would expect for the base particles in a regular BCC lattice site (Fig.~S6C blue). 

\subsection{Figure S7: Clustering of interstitial dopants}

Clustering occurs across the whole range of volume fractions for dynamic base crystals (Fig.~S7A-D top row), whereas it remains completely absent for the fixed base crystals. Clusters are identified using the DBSCAN algorithm as developed by Ester et al.[40] and embedded in the python library scikit-learn.  We consider particles with a maximum nearest neighbour distance of $1.5a$ to be in a single cluster, where $a$ is the lattice constant and the minimum cluster size is two. The heterogeneity in local cluster distribution is also reflected in the single-particle mean-squared displacements of the dopants. To this end we plot the distribution of $\langle \Delta r^{2}(\tau)\rangle$ values for all dopants in the simulation system (Fig.~S7A-D middle row). The distribution of MSD curves is highly heterogeneous, with different populations of dopant particles with similar MSD curves visible as thick bundles. To more clearly illustrate the heterogeneity of the MSD distribution we plot a distribution of $\langle \Delta r^{2}(\tau/\tau_b = 5 \cdot 10^2)\rangle$ (Fig.~S7A-D bottom row), which indeed shows multiple populations, which we attribute to singlets and clustered dopants.

\subsection{Figure S8: Heterogeneous long time dynamics.}
To investigate the cause of dynamical heterogeneity on long time scales we calculate the fraction of time a particle is present in a cluster. This fraction is plotted against $\langle \Delta r^2 \rangle$ at $\tau / \tau_b = 5 \cdot 10^2$ for every particle. 
This shows that at both high and low volume fractions there are two groups: clustered particles and singlets (Fig.~S8A \& B). Singlets only show low $\langle \Delta r^2 \rangle$ values, indicating strong localisation of singlets, while clustered particles show a broad range in these values. It is clear that at low volume fractions of the base crystal, where the elastic interactions responsible for clustering are weaker, the exchange of particles between singles and clusters is more dynamic, thus reducing the average time a particle resides in a clustered configuration.

\subsection{Figure S9: Cluster dynamics}

We start by looking at the fraction of total dopant particles in the system that are part of a cluster as a function of volume fraction $\phi$ (Fig.~S9A). The difference between the two states of the base crystal (static or dynamic) is striking. For the dynamic case we see that most of the particles form part of cluster and this ratio doesn't change significantly with $\phi$ (Fig.~S9A blue circles). For the fixed base crystal, the number of particles forming part of a cluster of $N \geq 2$ is very low and drops off as $\phi$ increases (Fig.~S9A red triangles). This seems to suggest that the clusters formed in the static case consist mostly of dimers due to the random thermal distribution of dopants across the lattice. 

Over time the fraction of particles present in a clustered, versus a singlet, state is constant (Fig.~S9D-F). For the dynamic crystal we observe small fluctuations in the fraction of clustered dopants around a mean value, indicative of the dynamic equilibrium between clusters and singlets. For the static cases these fluctuations are larger, as a result of the fact that these clusters are not formed by an effective medium attraction, driven by a minimization of the lattice strain but simply by random collisions of dopant particles as they diffuse across the lattice.

\subsection{Movie S1: Cluster dynamics $\phi = 0.07$}
To illustrate the dynamic behaviour of the dopant clusters in time and the continuous association and dissociation of dopants we reconstructured a movie of all dopant particles with those in a cluster, $S_C >= 4$, rendered larger to aid visualization. The movie spans a simulation time of $2.2 \cdot 10^3~\tau_b$. 

\subsection{Movie S2: Cluster dynamics $\phi = 0.12$}
Same as Movie S1 for $\phi = 0.12$.

\onecolumngrid

\begin{figure}%[tbhp]
\centering
\includegraphics[width=\linewidth]{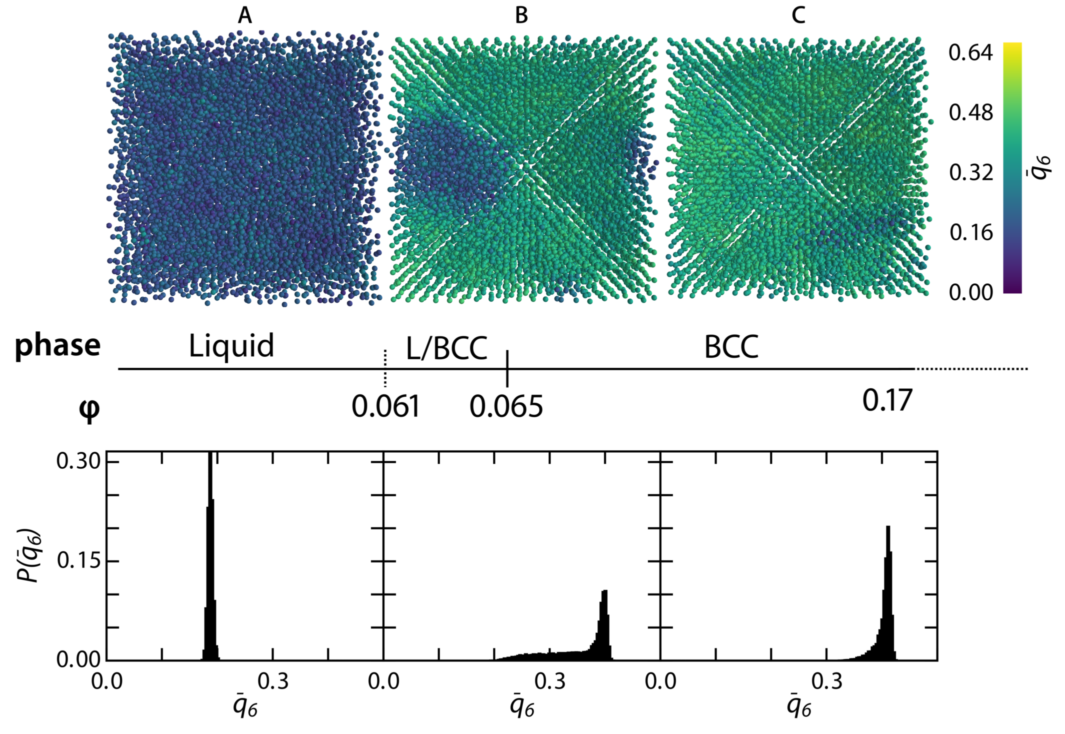}
\caption[Phase diagram and phase identification.]{Top: Base particles colored according to their $\bar{q}_{6}$ bond parameter. Middle: the phase diagram for our simulation system. Bottom: $\bar{q}_{6}$ histograms. (A) The liquid phase at $\phi = 0.060$. (B) Phase coexistence of liquid at the BCC phase at $\phi = 0.063$. (C) A BCC phase at $\phi = 0.0656$. Histograms are an average over 90 snapshots separated by $\Delta t = 2 \cdot 10^1$ $\tau_{b}$}
\label{fig:figureS1}
\end{figure}

\begin{figure}%[tbhp]
\centering
\includegraphics[width=\linewidth]{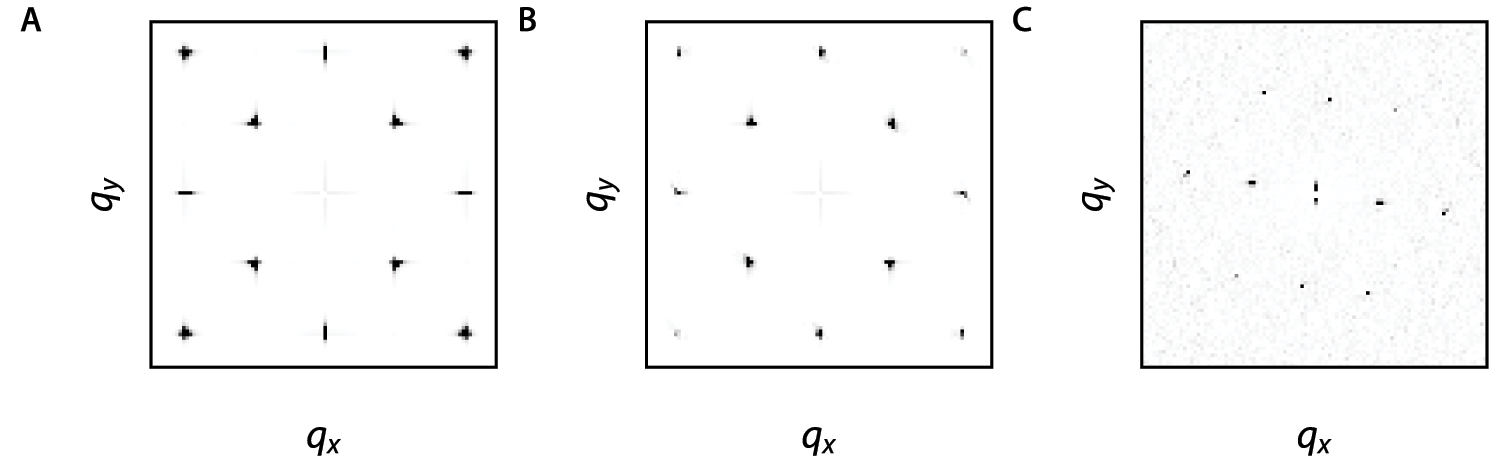}
\caption[Structure factor of static and dynamic base crystal]{$S(q_{x},q_{y})$ of the base particles of (A) a static base crystal with $\phi=0.07$, (B) a dynamic base crystal with $\phi=0.07$ and (C) experimental BCC crystal with $\phi=0.14$.}
\label{fig:figureS2}
\end{figure}

\begin{figure}%[tbhp]
\centering
\includegraphics[width=.66\linewidth]{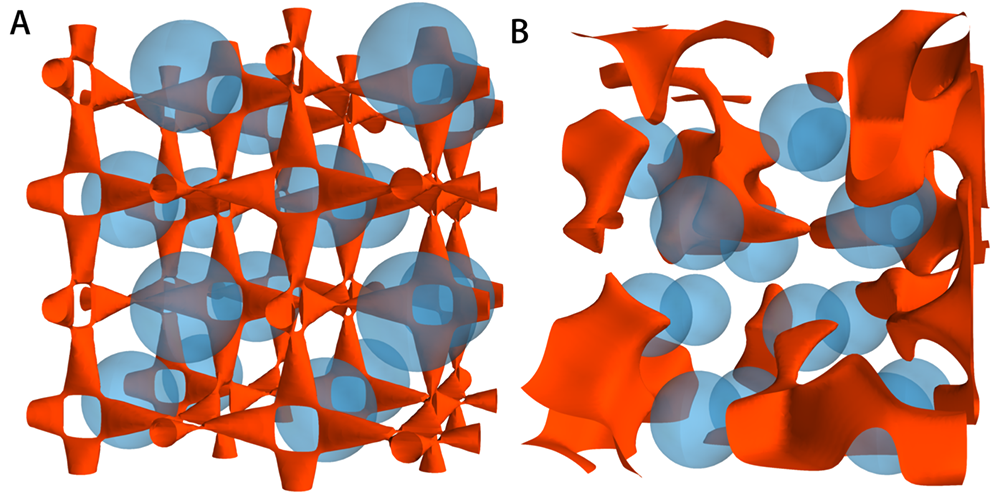}
\caption[Break up of the three dimensional percolated network]{Base particles (blue) with an isosurface (orange) plotted of the potential field felt by a dopant particle. (A) Static base crystal with isosurface at $\Delta U = 3.5 k_{B}T$ and (B) potential field of a snapshot at $t = 2 \cdot 10^3 \sigma_b$ with an isosurface at $\Delta U = 42 k_{B}T$}
\label{fig:figureS3}
\end{figure}

\begin{figure}%[tbhp]
\centering
\includegraphics[width=.66\linewidth]{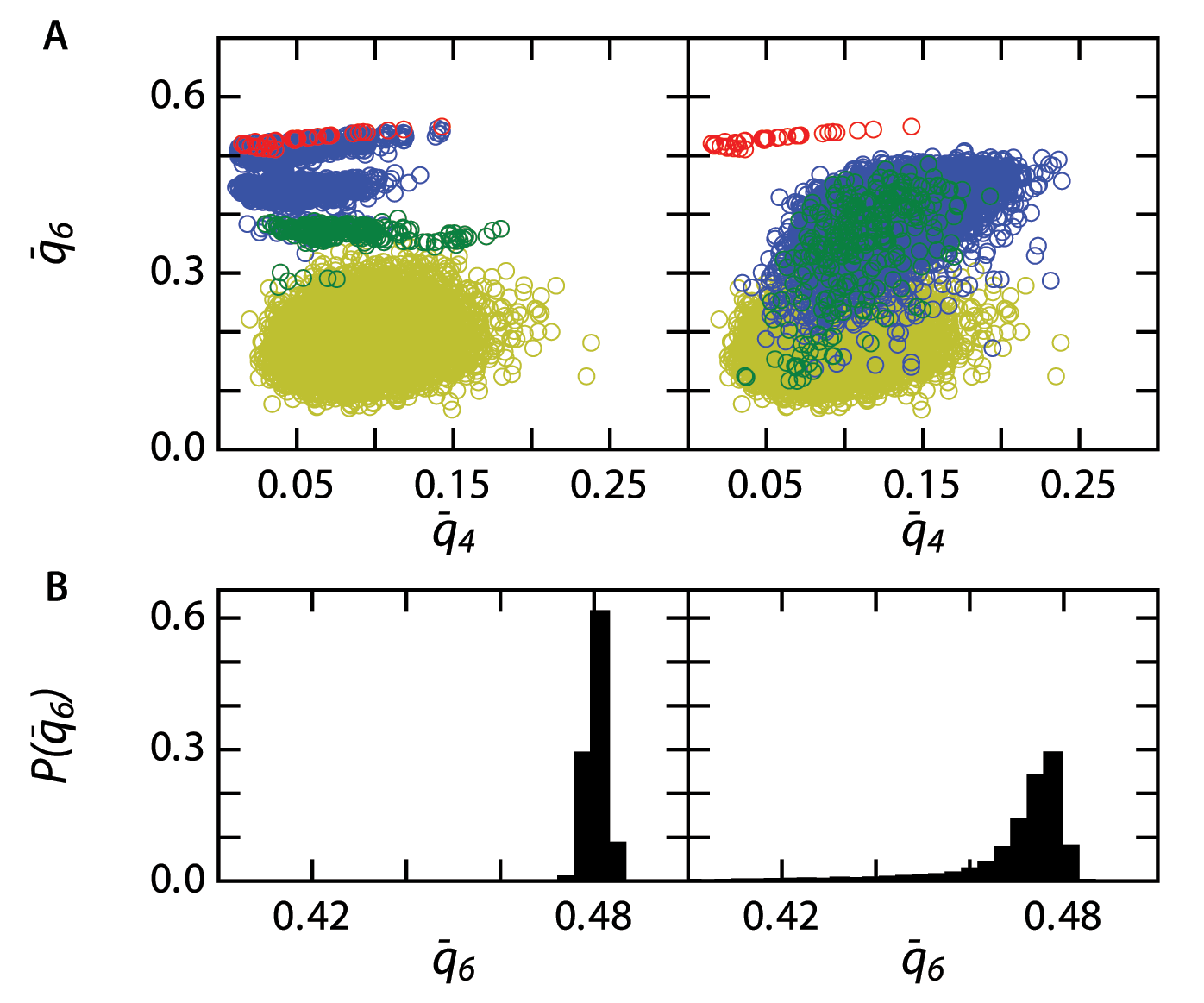}
\caption[Influence of dopants on the structure of the base crystal]{The $\bar{q}_6$ versus $\bar{q}_4$ for every particle in a snapshot of the system at $t= 2 \cdot 10^3$ $\tau_{b}$. for (A) a static base crystal at $\phi = 0.07$ and (B) a dynamic base crystal at $\phi = 0.07$. (C) The probability distribution for $\bar{q}_6$ in a system without dopant particles (left) and $N_d:N =$~1:47 (right). This is an average over 90 snapshots separated by $\Delta t = 2 \cdot 10^1$ $\tau_{b}$}
\label{fig:figureS4}
\end{figure}

\begin{figure}%[tbhp]
\centering
\includegraphics[width=.66\linewidth]{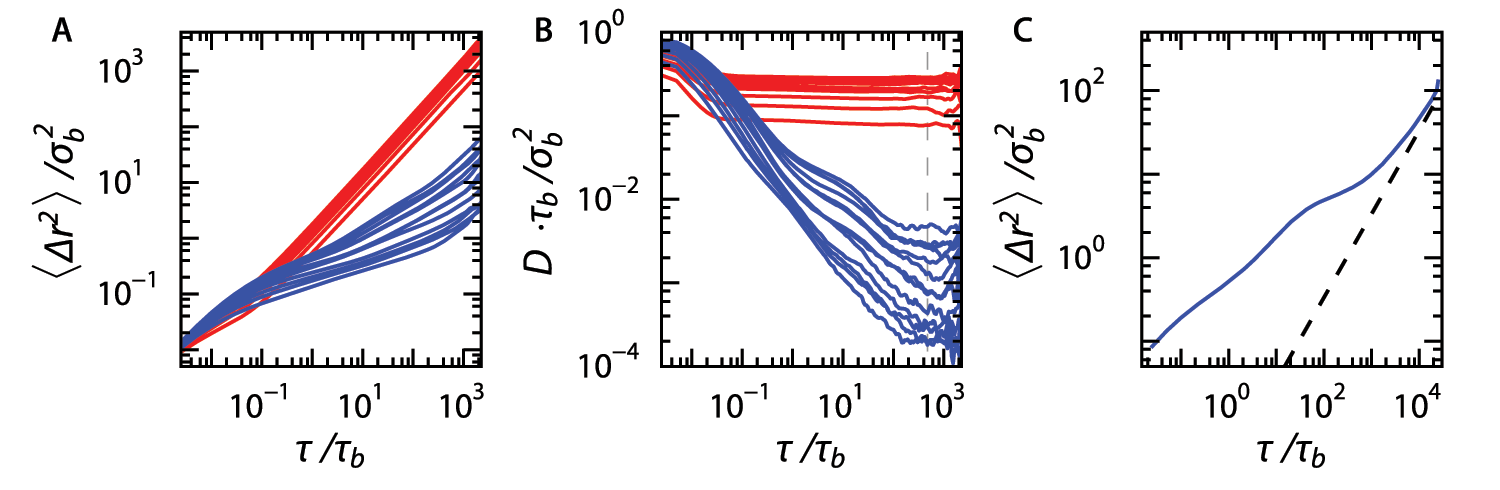}
\caption[Mean squared displacement and diffusion.]{Ensemble averaged mean-squared displacements for all volume fractions, $\phi=$~0.0656, 0.066, 0.067, 0.070, 0.071, 0.075, 0.08, 0.09, 0.10, 0.12, 0.15, and 0.20. (A) $\langle \Delta r^{2}(\tau)\rangle$ of the dopant particles for a static base crystal (red) and a dynamic base crystal (blue). Volume fraction increases from bottom to top (B) The diffusion coefficient $D(\tau)$ calculated from the gradient of $\langle\Delta r^{2}(\tau)\rangle $ for the static base crystal (red), the dynamic base crystal with $N_d:N =$~1:47 (blue). $\phi$ decreases from bottom to top. (C) Ensemble averaged mean-squared displacements for a longer simulations; the extra order of magnitude in simulation time clearly shows the upturn towards diffusive behavior; dashed line has a slope of one.}
\label{fig:figureS5}
\end{figure}

\begin{figure}%[tbhp]
\centering
\includegraphics[width=\linewidth]{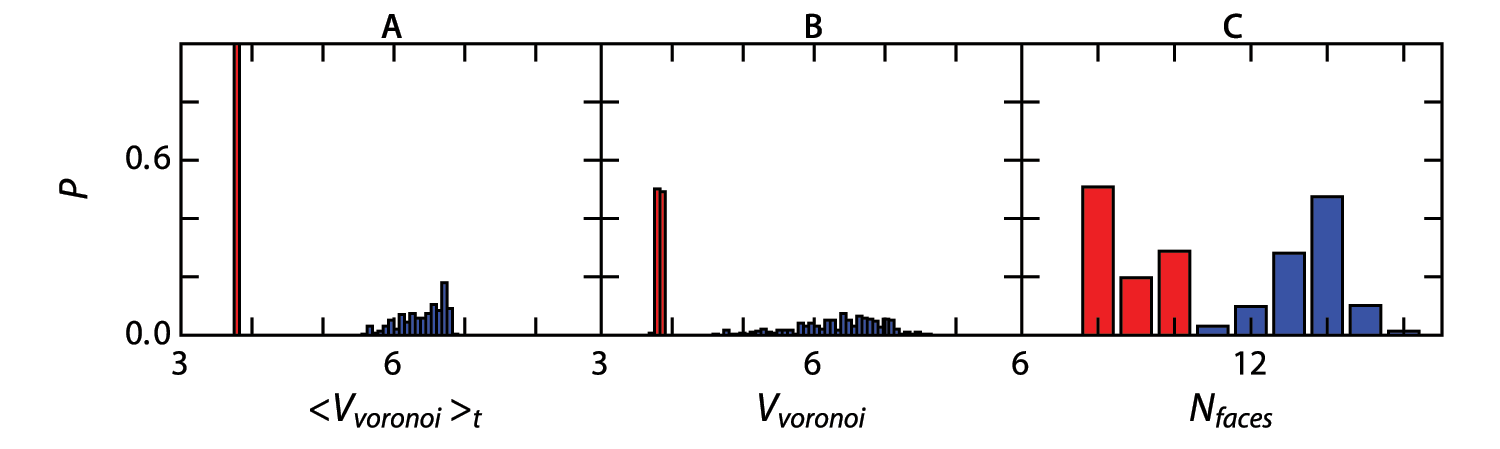}
\caption[Voronoi analysis]{Results from Voronoi analysis on dopant particles at $\phi = 0.07$ for a static base crystal (red) and a dynamic base crystal (blue) (A) The average voronoi volume, (B) the voronoi volume at a snapshot at $t= 2 \cdot 10^3$ $\tau_{b}$ and (C) the number of faces at $t= 2 \cdot 10^3$ $\tau_{b}$.}
\label{fig:figureS6}
\end{figure}

\begin{figure}%[tbhp]
\centering
\includegraphics[width=\linewidth]{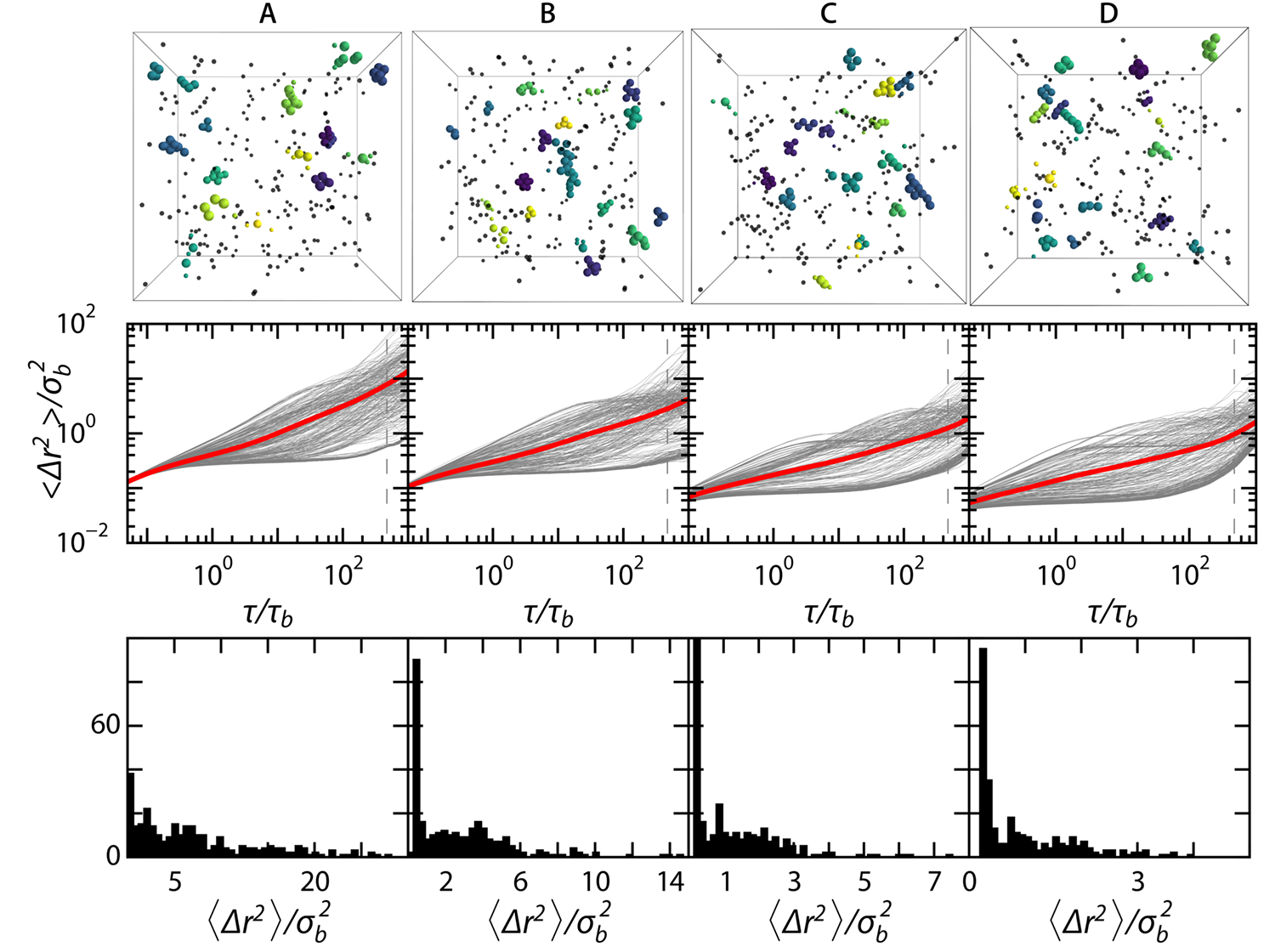}
\caption[Cluster formation and heterogeneity.]{Cluster analysis at different volume fractions: (A) $\phi = 0.07$, (B) $\phi = 0.09$, (C) $\phi = 0.15$ and (D) $\phi = 0.20$. Systems all have $N_d:N =$~1:47 Top: Snapshots of cluster formation at $t= 2 \cdot 10^3$ $\tau_{b}$, only the clusters with $S_C >= 4$ are highlighted for clarity. Middle: $\langle \Delta r^{2}(\tau)\rangle$ curves per particle (grey) and the ensemble average (red). Bottom: Histograms of the $\langle \Delta r^{2}\rangle$ at $\tau = 5 \cdot 10^2$ over 50 bins. }
\label{fig:figureS7}
\end{figure}

\begin{figure}%[tbhp]
\centering
\includegraphics[width=.66\linewidth]{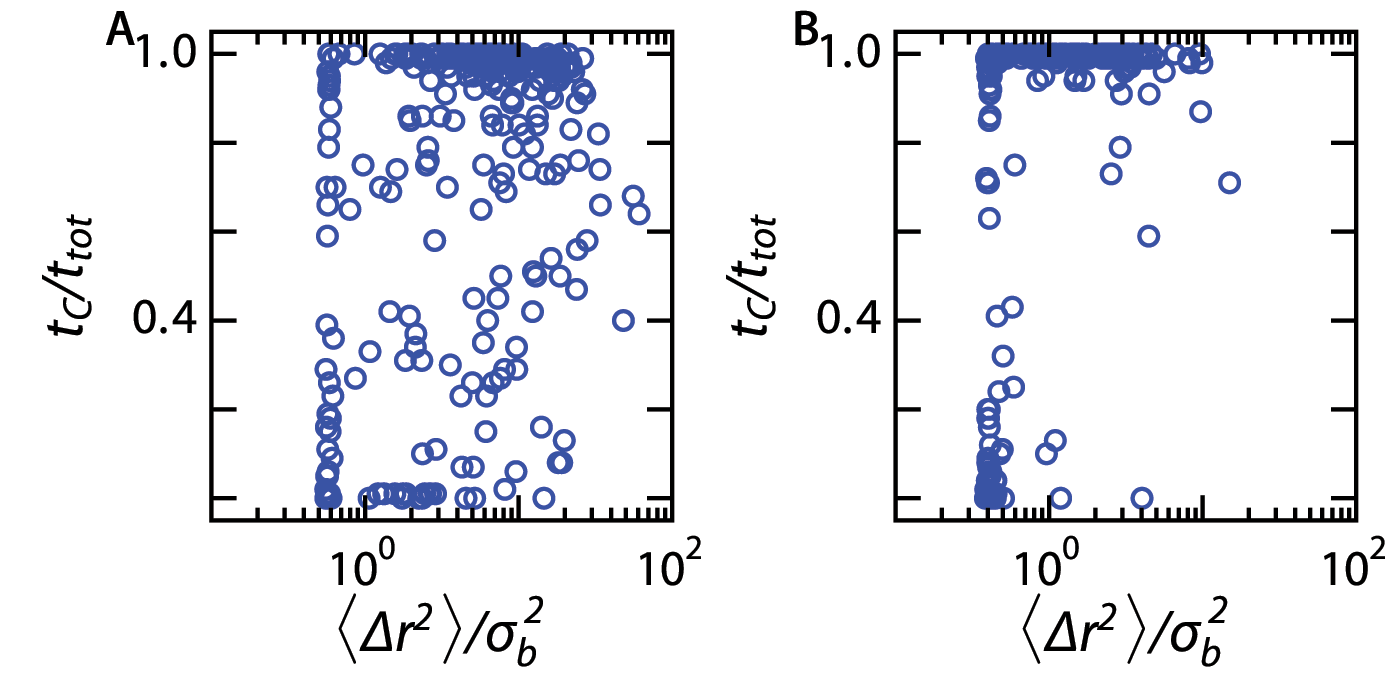}
\caption[Heterogenous long time dynamics]{$\langle \Delta r^2\rangle$ at $t=5\cdot10^2$ versus the fraction of simulation time that a particle is part of a cluster. Plotted for volume fractions: (A) $\phi=0.07$ and (B) $\phi=0.12$.}
\label{fig:figureS8}
\end{figure}

\begin{figure}%[tbhp]
\centering
\includegraphics[width=\linewidth]{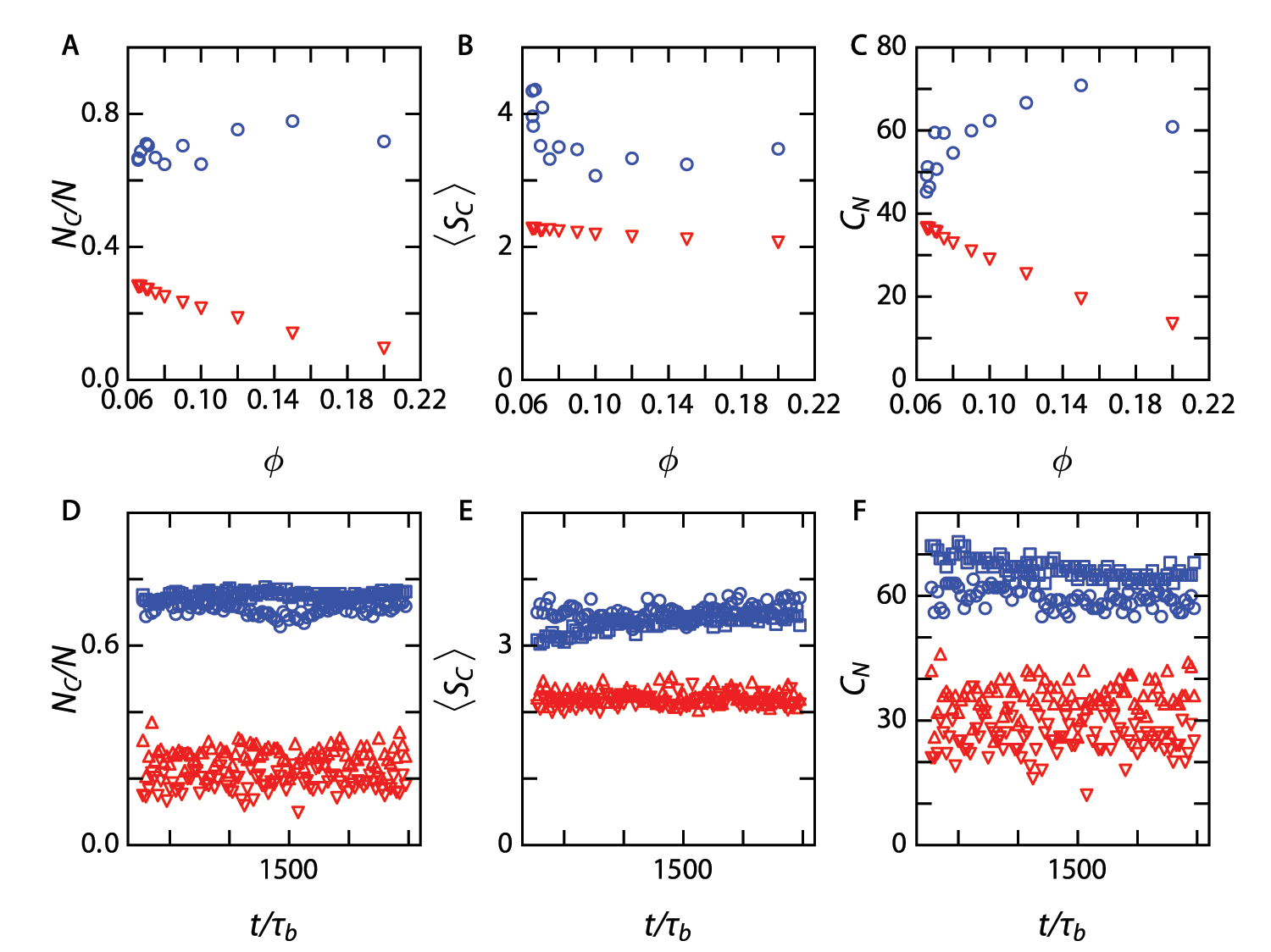}
\caption[Cluster dynamics]{(A) The average fraction of clustered particles $N_C/N$ against $\phi$. (B) Averaged cluster size $S_C$ as a function of $phi$ and (C) average amount of clusters $C_N$ against $\phi$. For $N_d:N =$~1:47 in a static base crystal (red triangles) and in a dynamic base crystal (blue circles) (D-F) Behaviour of $N_C/N$, $\langle S_C \rangle$, and $C_N$ during the simulation at two different volume fractions $\phi = 0.07$ (blue circles and red downward pointing triangles) $\phi = 0.12$ (blue squares and red upward pointing triangles).}
\label{fig:figureS9}
\end{figure}

\end{document}